\documentclass[12pt,preprint]{aastex}

\def\be{\begin{equation}}
\def\ee{\end{equation}}

\usepackage{natbib}

\begin{document}

\title{HIGH ANGULAR RESOLUTION
SUNYAEV ZEL'DOVICH
OBSERVATIONS: THE CASE OF
MISTRAL}

\author{E. S. Battistelli$^1$$^,$$^*$, E. Barbavara$^1$, P. de Bernardis$^1$, F. Cacciotti$^1$, V. Capalbo$^1$, E. Carretti$^2$, F.
Columbro$^1$, A. Coppolecchia$^1$, A. Cruciani$^3$, G. D'Alessandro$^1$, M. De
Petris$^1$, F. Govoni$^4$, G. Isopi$^1$, L. Lamagna$^1$, P. Marongiu$^4$, S. Masi$^1$, L.
Mele$^1$, E. Molinari$^4$, M. Murgia$^4$, A. Navarrini$^4$, A. Orlati$^2$, A. Paiella$^1$, G.
Pettinari$^5$, F. Piacentini$^1$,T. Pisanu$^4$, S. Poppi$^4$, G. Presta$^1$, F. Radiconi$^1$}

\affil{$^{1}$Dipartimento di Fisica, Sapienza Universit\`a di Roma - Piazzale Aldo Moro, 5, I-00185 Rome, Italy}

\affil{$^{2}$INAF - Istituto di Radioastronomia - Via P. Gobetti, 101 - I-40129 Bologna, Italy}

\affil{$^{3}$INFN - Sezione di Roma, Piazzale Aldo Moro 5 - I-00185, Rome, Italy}

\affil{$^{4}$INAF - Osservatorio Astronomico di Cagliari, Via della Scienza 5 - I-09047 Selargius (CA), Italy}

\affil{$^{5}$Istituto di Fotonica e Nanotecnologie - CNR, Via Cineto Romano 42, I-00156 Rome, Italy}

\altaffiltext{*}{e-mail: elia.battistelli@roma1.infn.it}

\begin{abstract}
The MIllimeter Sardinia radio Telescope Receiver based on Array of Lumped elements kids, MISTRAL, is a millimetric ($\simeq 90GHz$) multipixel camera being built for the Sardinia Radio Telescope. It is going to be a facility instrument and will sample the sky with 12 arcsec angular resolution, 4 arcmin field of view, through 408 Kinetic Inductance Detectors (KIDs). The construction and the beginning of commissioning is planned to be in 2022. MISTRAL will allow the scientific community to propose a wide variety of scientific cases including protoplanetary discs study, star forming regions, galaxies radial profiles, and high angular resolution measurements of the Sunyaev Zel'dovich (SZ) effect with the investigation of the morphology of galaxy cluster and the search for the Cosmic Web.
\end{abstract}


\section{Introduction}\label{aba:sec1}

High angular resolution millimetric observation are key to understand a wide variety of scientific cases. Among others, the interaction of Cosmic Microwave Background (CMB) photons with the hot electron gas in galaxy clusters and surrounding medium (the Sunyaev Zel'dovich, SZ effect) promises to study galaxy clusters and their deviation from relaxed behaviour. This include high angular resolution measurements of the SZ effect with the investigation of the morphology of galaxy cluster and the search for the Cosmic Web. Also, protoplanetary discs study, star forming regions, and galaxies radial profiles are among the important scientific cases one could achieve with high angular resolution millimetric observation.

Only a handful of instruments are capable of measuring, for instance, galaxy clusters with high enough angular resolution to resolve them. Among them, we mention MUSTANG2 at the Green Bank Telescope (GBT\footnote{\url{https://greenbankobservatory.org/science/telescopes/gbt/}}) and NIKA2 at IRAM telescope\footnote{\url{https://www.iram-institute.org/EN/30-meter-telescope.php}}. MUSTANG2 \cite{dickermustang2} is a millimetric cryogenic camera installed at the 100m GBT. It is operating in the frequency range between 75GHz and 105GHz and samples 4 arcmin f.o.v. in the sky with 223 microstrip-coupled  Transition Edge Sensors (TESs). The resulted diffraction limited angular resolution is of the order of 9 arcsec. A frequency domain multiplexing technique is used to read MUSTANG2 detectors through flux-ramp modulation technique. 

Another receiver which addresses similar science goals, although it takes data at higher frequencies, is NIKA2 \cite{Catalano2018, Adam2018, Perotto2020}. NIKA2 is a 2 channel millimetric camera measuring at 150GHz (one array) and 260GHz (two arrays). It observes from the 30m IRAM telescope with a field of view (f.o.v.) of 6.5 arcmin while the angular resolution is around 18 arcsec at 150GHz and around 10 arcsec at 260GHz. NIKA2 hosts 3 arrays of Kinetic Inductance Detectors (KIDs), each of around 2000 pixels, cooled down at 100mK with a dilution refrigerator. Both MUSTANG2 and NIKA2 are very active in the field of millimetric astronomy especially for the high angular resolution detection of the SZ effect.

A third instrument being prepared for such scientific cases is the MIllimeter Sardinia radio Telescope Receiver based on Array of Lumped elements kids, MISTRAL, to be fielded at the 64m Sardinia Radio Telescope (SRT\footnote{\url{http://www.srt.inaf.it/}}). MISTRAL will use an array of 408 KIDs with a f.o.v. of 4 arcmin and an angular resolution of around 12 arcsec. It is a cryogenic instrument with detectors cooled down at 270mK and with frequency domain multiplexing, ROACH2 based read-out.

\section{Sardinia Radio Telescope and its upgrade}

\subsection{Sardinia Radio Telescope}
\begin{figure}[h]
    \centering
    \includegraphics[width=\textwidth]{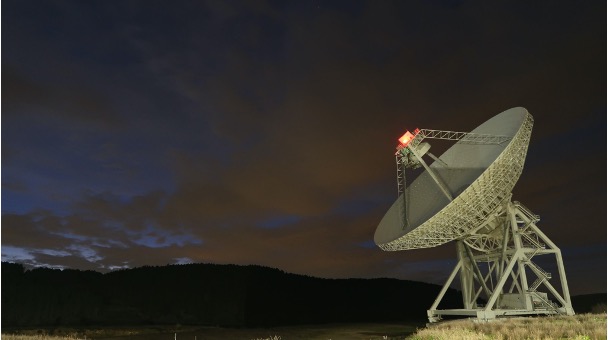}
    \caption{The 64-m Sardinia Radio Telescope (image credits F. Guidi).}
    \label{fig:SRT}
\end{figure}

The Sardinia Radio Telescope (SRT - Lat. 39.4930 N; -Long. 9.2451 E), is a Gregorian configured fully steerable 64m primary mirror radio telescope which can work from 300MHz and 116GHz\cite{Bolli_2015_commissioning} (see figure \ref{fig:SRT}). It is a multipurpose instrument operated in either single dish or Very Long Baseline Interferometer mode and managed by the Italian Astrophysics National Institute (INAF). The telescope manufacturing started in 2003 and was completed in August 2012 when the technical commissioning started. The scientific exploitation of the SRT started on 2016 with an Early Science Program, while regular proposals started in 2018. 

SRT has a f/0.33 primary focus and f/2.34 secondary focus allowing to host the low frequency receivers ($\nu<2GHz$) in the primary focus, and high frequency receivers in the Gregorian room. The 64-m primary mirror (M1) is composed of 1008 electromechanically controlled aluminum elements by actuators. The secondary mirror (M2) is a 7.9m mirror composed of 49 aluminum elements adjustable for focus operations. Both M1 and M2 are shaped to minimize spillover and the standing waves between the receiver and the subreflector\cite{Bolli_2015_commissioning}. 

\subsection{Observational site}
The location of the SRT is in the island of Sardinia, $\simeq$50km north of Cagliari, at 600m a.s.l.. It is a region suited for radio-observations and the telescope was designed to be operating up to 116GHz. Estimation of sky opacity, based on recorded dedicated radiometer data, reports\footnote{\url{http://hdl.handle.net/20.500.12386/28787}}  an opacity $\tau<0.15Np$ (50th percentile) at 93GHz during the winter nights. The average precipitable water vapour (PWV) in the same conditions is mainly 8mm. For comparison, the GBT reports $\tau<0.125$ (50th percentile) at 86GHz, and an average $PWV<9mm$ (50th percentile)\footnote{\url{https://www.gb.nrao.edu/mustang/wx.shtml}}. The above estimation is confirmed by radiosondes profiles taken at Cagliari airport and scaled for the SRT site. This shown $PWV<11mm$ (50th precentile) and opacities $<0.2Np$ (50th percentile) at 100GHz \cite{Nasir_2011}.

\subsection{PON upgrade}
In 2018, a National Operational Program (PON) grant was assigned to INAF with the aim to exploit to the full the SRT capability to reach mm wavelenghts up to 116GHz\cite{Govoni2021}. One of the Working Packages includes a metrological system to measure with high precision deformations of the SRT due to gravitational and thermal effects; by means of the active surface it will be possible to correct the deformations of the main mirror allowing the surface roughness needed for observations  up to 116 GHz. Also, the metrological system will allow high precision pointing measuring and correcting  the  subreflector displacement   from the optical axis  together with the deformation of the whole structure.

In addition an upgrade of the SRT receiver fleet is foreseen including an extension to higher frequencies up to 116GHz, thus exploiting to the full its optical surface. This grant was divided in 9 working packages which include new receivers like a 19 feeds, double polarization Q-band heterodyne receiver, a triple band (K, Q, W) coherent receiver, a 16 double polarization W-band heterodyne receiver, and a W-band, total power, 408 pixels bolometric\footnote{Actually MISTRAL will implement Kinetic Inductance Detectors which are not properly bolometers.} receiver: MISTRAL. 

\section{MISTRAL}

\subsection{Introduction}

MISTRAL is a facility instrument to be installed at the SRT in 2022. It operates in an atmospheric window in the frequency range 78-103GHz, namely W-band, that is interesting for a number of scientific reasons, and most importantly it provides low optical depth and allows high efficiency observations. It will be installed in the Gregorian room of the SRT and as such, it needs to meet several requirements and limitations. These include:

\begin{itemize}
\item 250kg maximum weight;
\item 700mm x 700mm x 2400mm (h) maximum occupation. In addition, the top part should not interfere with the Gregorian room structure;
\item adeguate Radio Frequency shielding;
\item ability to work from 30$^{\circ}$ elevation to 80$^{\circ}$ and the need not to be damaged in the range 0-90$^{\circ}$ elevation;
\item long ($\simeq$100m) long helium lines for the Pulse Tube head-compressor connection;
\item remote operations;
\item MISTRAL will be installed on the Gregorian room turret (a rotating structure which hosts MISTRAL an all other receivers) and needs to be agile enough to be inserted in the SRT focus or removed through rotation of the turret itself (see figure \ref{fig:cryo}).
\end{itemize}

\subsection{Cryostat}

MISTRAL consists of a cryostat, being constructed by QMC instruments\footnote{\url{http://www.terahertz.co.uk/qmc-instruments-ltd}}, with a Pulse Tube cryocooler, an He-10 sorption refrigerator, custom optics and detectors. We use a Sumitomo\footnote{\url{https://www.shicryogenics.com/}} RP-182B2S-F100H Pulse Tube (PT) cryocooler. From Sumitomo user manuals, PT heads work properly in vertical position with inclination no larger than +/-20$^{\circ}$. In order to have the PT heads as close as possible to vertical position during observations, we have positioned the head with an inclination of 57.5$^{\circ}$ with respect to the focal plane. This allows a nominal observation elevation of 37.5-77.5$^{\circ}$. Nevertheless, we have tested higher inclinations and there is no impact on the cryogenics with inclination of +/-25$^{\circ}$ (resulting in elevation range of 32.5-82.5$^{\circ}$) with no degradation of the thermal performance.  MISTRAL has a 40$K$ shield, a 4.2$K$ shield and a sub-Kelvin stage. The PT is used to cool down the 40$K$ and 4.2$K$ shields and provides a cooling power of 36$W$ for the 40$K$ shield and 1.5$W$ for the 4.2$K$ shield.

\begin{figure}[h]
    \centering
    \includegraphics[width=140mm]{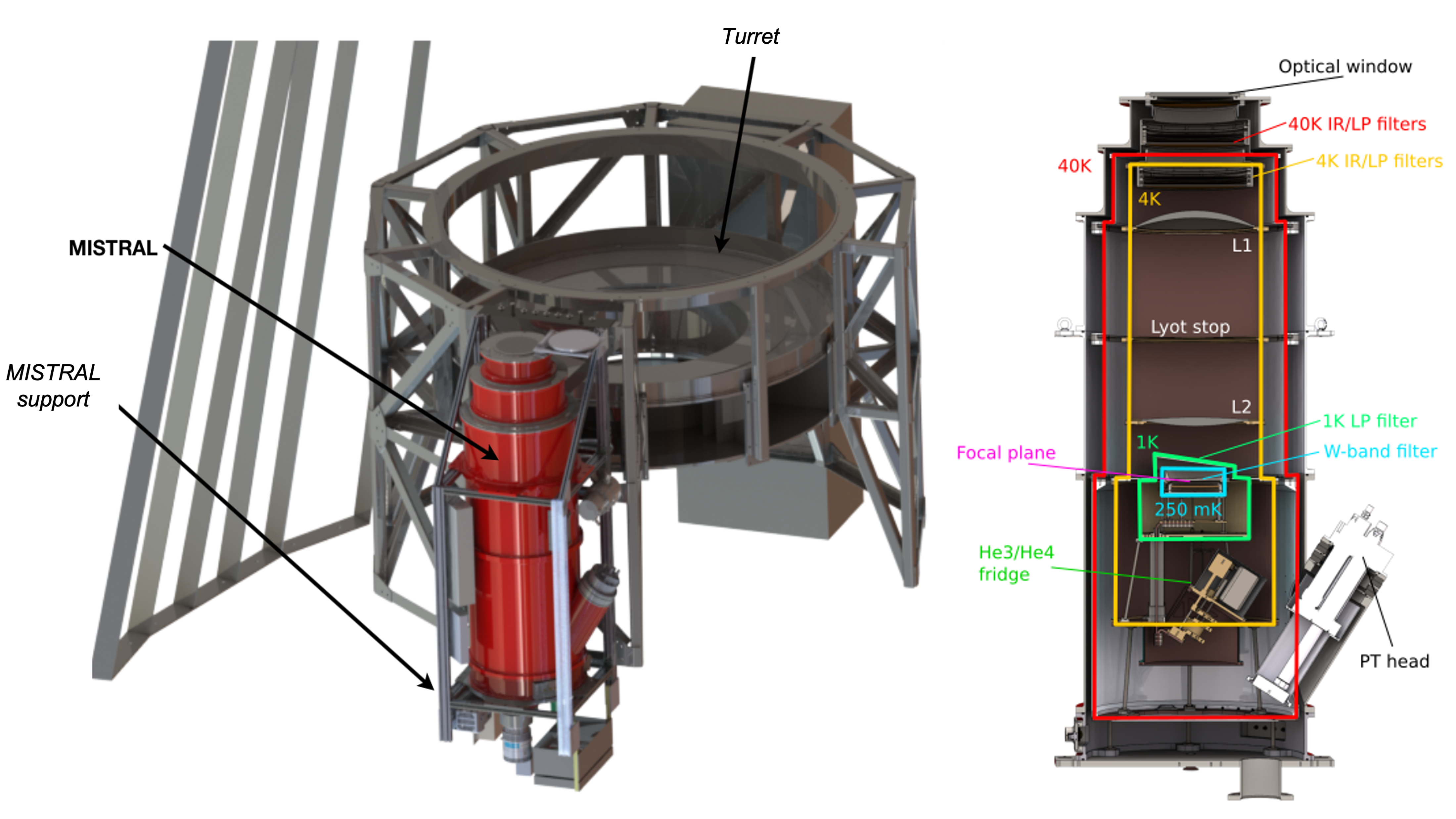}
    \caption{Left: SRT Turret. with MISTRAL installed. Right: a cut of MISTRAL cryostat highlighting the main parts.}
    \label{fig:cryo}
\end{figure}

The PT head is controlled by a PT compressor which cannot be inclined with the telescope elevation angle. For this reason, MISTRAL compressor will not be located at the Gregorian room, but rather in a compressor room at the base of the radio telescope, that can rotate with the azimuth angle but does not change inclination with the elevation angle. This requires about 100m of flexible lines for the compressor-to-PT head connection. Sumitomo (as well as other companies) report 20m maximum length for this operations so on-site tests were performed in order to verify if a PT can keep the 4.2$K$ base temperature and with which dissipated power\cite{coppolecchia_LTD19}. We made these tests using a Sumitomo RP-082B2S cryocooler, which is similar to the one chosen for MISTRAL but with lower cooling power (i.e. 0.9$W$ at 4.2$K$). It was shown that this cryocooler was able to hold a base temperature of 4.2$K$ with a power load of 0.6W with 119$m$ of flexible lines. Since the power dissipation requirement for MISTRAL is of the order of 0.6$W$, the 1.5$W$ PT will be able to satisfy the requirement with some margin. In addition, a vibration dumper, and a remote PT valve, will allow to reduce vibration into the detectors.

Attached to the 40$K$ and the 4.2$K$ stages, it will be applied radiation shields as well as radiation filters allowing to block thermal input of the sub-K stages of MISTRAL. The sub-K fridge is a Chase\footnote{\url{https://www.chasecryogenics.com/}} He$^3$/He$^4$ sorption refrigerator (Twin GL10 sorption fridge) allowing a nominal cooling power as in the following table:

\begin{center}
\begin{tabular}{ |c|c|c|c| } 
 \hline
  Stage & temperature no load & temperature loaded & Load applied  \\ 
\hline
 He$^{4}$ & 0.840$K$ & 0.913$K$ & 150$\mu W$  \\ 
 He$^{3}$ &0.299$K$ & 0.332$K$ & 30$\mu W$ \\ 
 Ultra cold He$^{3}$ &0.195$K$ & 0.251$K$ & 20$\mu W$ \\ 
 \hline
 \end{tabular}
\end{center}

The total room to be cooled down is of the order of 1.5$m^{3}$ and the total weight of MISTRAL is around 250$kg$. The top part of the cryostat has been shaped in such a way it does not interfere with the Gregorian room walls while rotating in and out of the focus position (see figure \ref{fig:cryo}).

\subsection{Optics and quasi-optics}

The optical design of MISTRAL includes a set of radiation quasi-optical filters, anchored at the different thermal stages of the cryostat, an Anti Reflection Coated (ARC) Ultra High Molecular Weight (UHMW) polyetilene window, and two ARC Silicon lenses able to image the Gregorian Focus on the array of detectors. Detectors are coupled to radiation through open space (filled array) so cryogenic cold stop, placed in between the two lenses, is needed.

Quasi-optical filters are a combination of metal mesh filters, thin IR filters, sub-mm low pass filters (LPE),  and a final 78-103GHz Band pass filter produced at QMC instruments\footnote{\url{http://www.terahertz.co.uk/qmc-instruments-ltd}}. The final configuration still needs to be settled down as it depends on cryogenic tests and performance that still need to be carried on. Nevertheless, the filter chain will include:

\begin{itemize}
\item 300$K$: ARC UHMW window of 190mm outer diameter, plus two blocking thermal filters; 
\item 40$K$: two additional thermal filters plus a Low Pass Edge filter;
\item 4$K$: two more thermal filters (IR5 and IR6) and another Low Pass filter;
\item 1$K$: Low Pass filters;
\item 0.350 $K$: Low Pass filters;
\item 0.250 $K$: Band Pass Filter centered at 90GHz with a bandwidth either of 30GHz or of 20GHz (to be defined).
\end{itemize} 

Some characteristics of the SRT optical configuration (which were adopted to make the SRT suitable for a wide range of applications) limit the Diffraction Limited Field Of View (DLFOV) usable by the mm camera. As mentioned, we plan to implement a naked Kinetic Inductance Detectors (KIDs) array ( a filled array) optically coupled to the antenna by a relay optics. This is the preferred solution for satisfying the following requirements:

\begin{itemize}
\item to possibly rescale the Gregorian telescope focal plane;
\item to insert a cold aperture stop;
\item to provide a gaussian wavelength independent beam telescope configuration;
\item to provide a telecentric optics.
\end{itemize}

The first requirement allow us to tune the size of the KID array preserving the telescope DLFOV. The second condition allows us to introduce an aperture conjugated to the secondary mirror, reduces the spillover contribution due to subreflector and/or primary mirror, ensuring the highest edge taper in a cold aperture. The third condition enables a magnification of the relay optics and the output beam waist location wavelength-independent, at least inside the expected bandwidth, which is a considerable innovation point. The fourth requirement ensures a homogeneous illumination of the array, apart from a possible aberrations impact, suitable for the solution of a naked array maintaining an adequate Strehl Ratio.

Silicon is a suitable material for lens fabrication at millimetri wavelenghts due to its high dielectric constant, low loss tangent, and high thermal conductivities. The two Silicon lenses allow to report 4 arcmin of the SRT focus onto the array of 408 KIDs. They are anti-reflection coated with Rogers RO3003. Their diameter is 290mm and 240mm respectively while the aperture cold stop diameter of 125mm. All the lenses+cold stop system is kept at 4K. The in band average simulations report excellent values with a Strehl Ratio from 0.97 to 0.91 for on-axis and F. O. V. edge positions. Analogously, the FWHM is 12.2 arcsec on axis, and 12.7 arcsec at 45mm off axis.

\subsection{Detectors}

MISTRAL will take advantage of the high sensitivity and the capability of Frequency Domain Multiplexing of KIDs cryogenic detectors. KIDs are superconductive detectors where millimetric radiation with higher energy with respect to the Cooper pair binding energy, can break Cooper pairs producing a change in the population densities and thus in the kinetic inductance. In fact, the inductance $L$ of a thin superconductor is dominated by the kinetic inductance $L_k$, which depends on the Cooper pair density.

If we integrate this inductance into a resonating RLC circuit, with a well defined resonant frequency, we will see a change in the impedance of the circuit when the inductance is illuminated. This change in impedance can be measured by exciting the circuit with an RF tone at the resonant frequency and monitoring both amplitude and phase of the outcoming tone. The amplitude, and similarly the phase, will change because of the variation in $L_k$. The general effect is that a decrease in the Cooper pair density increases $L_k$, thus it lowers the resonant frequency. Furthermore, the increase of the quasi-particle density increases the internal dissipations, thus reducing the quality factor of the resonator.  KIDs are intrinsically easy to multiplex in the frequency domain because an array of resonators with different resonant frequencies can be simultaneously readout by sending a comb of Radio Frequency (RF) tones through a single feed-line.

\begin{figure}[h]
    \centering
    \includegraphics[width=140mm]{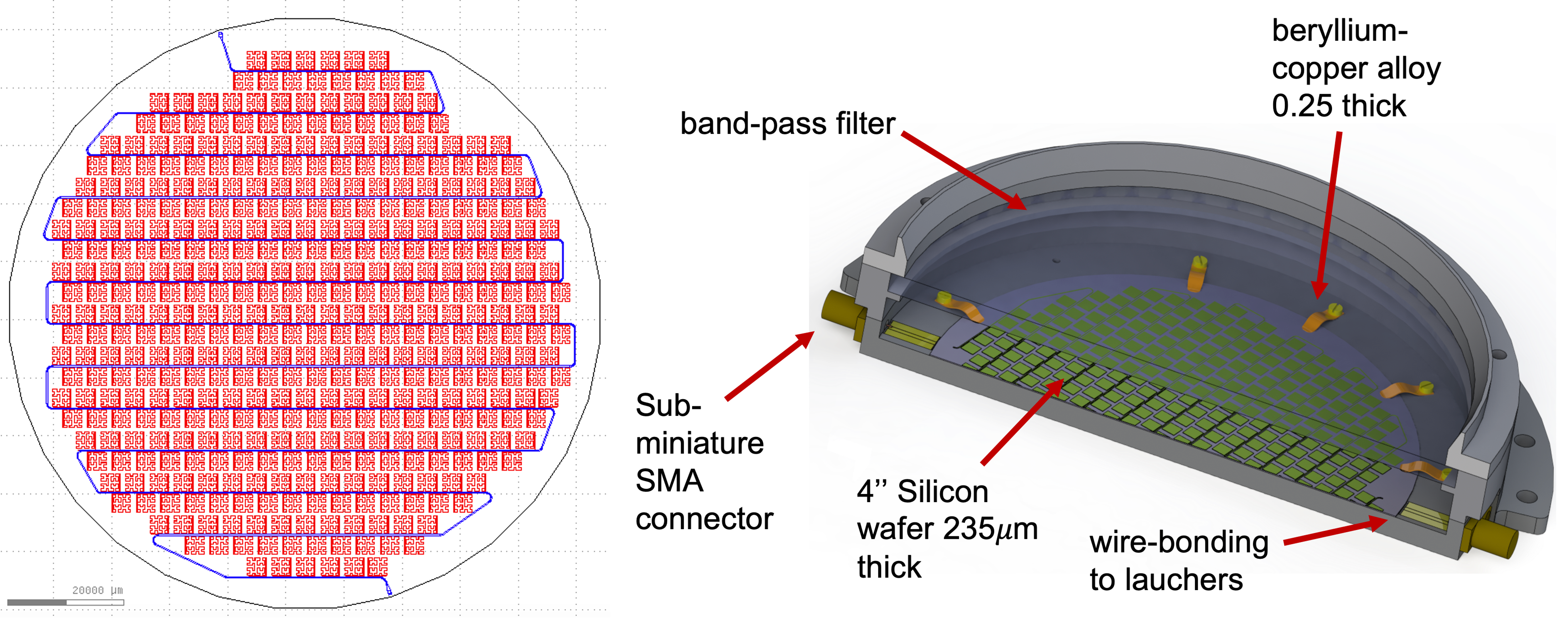}
    \caption{Left: a design of the focal plane of MISTRAL with its feedline\cite{paiella_LTD19}. Right: KIDs holder with the surrounding components.}
    \label{fig:KIDs}
\end{figure}

MISTRAL KIDs are Ti-Al bilayer 10 + 30 $nm$ with critical temperature Tc=945mK and are fabricated at CNR-IFN\footnote{\url{https://www.roma.ifn.cnr.it/}} \cite{Paiella2016,Coppolecchia2020}. They are front-illuminated 3rd order Hilbert crude absorber with backshort on the opposite side of the wafer. The detector array is composed of $408$ KIDs detectors. They are 3mm x 3mm each and are arranged on an equilateral triangle every 4.2mm on a 4 inches silicon wafer (see figure \ref{fig:KIDs}). They sample the focal plane with a FWHM angular spacing of 10.6 arcsec lower the each pixel angular resolution.

As mentioned, KIDs behave as high quality factor, Q, LC resonators. High values of Q allow to multiplex thousands of KIDs, with different frequencies, all coupled to the same feedline. KIDs are in fact intrinsically easy to multiplex in the frequency domain, because an array of resonators with different resonant frequencies can be simultaneously readout by sending a comb of RF tones through a single feed-line. We will use ROACH2 based Frequency Domain Multiplexing which was originally developed for the OLIMPO experiment\cite{Paiella_2019_olimpo}.

\subsection{Magnetic shield}

Another important part of MISTRAL is a 1-mm thick Cryoperm 10 $\mu$-metal shield, characterized by an high magnetic permeability ($\mu > 7000$), that acts as a shielding from spurious magnetic fields such as the geomagnetic field. When a KID, usually composed of a thin wire, is exposed to an external magnetic field, a shielding current will be induced. Consequentially, its kinetic inductance changes. The external magnetic field will increase the kinetic inductance of the circuit, thus increasing the total inductance, eventually reducing the quality factor of the resonator. This effect was observed on the MISTRAL detector array \cite{paiella_wolte14} and required a magnetic shield to be lower than the earth magnetic shield. 

\section{Sensitivity forecast and scanning strategy}

\subsection{Noise estimate}

A crucial step to predict the performance on the sky is estimating the expected noise which should affect our detectors. Electrical Noise Equivalent Power (NEP) should be compared with Photon noise NEP due to warm optics and atmosphere emission. The latter, should be considered in its emission part and turbulence part which will act as correlated noise.

Electrical NEP of MISTRAL detectors was measured to be temperature dependent. If we assume a final detectors temperature of 270mK we have average values of our test-bed array of detectors of $NEP_{avg}=3.5\times 10^{-16}W/\sqrt{Hz}$ with a maximum and a minimum range of  $NEP_{max}=7.4\times 10^{-16}W/\sqrt{Hz}$ and $NEP_{min}=1.6\times 10^{-16}W/\sqrt{Hz}$ respectively. These numbers need to be compared with estimated photon noise and turbulence noise from atmosphere. The photon noise part was estimated to be, within the 30GHz bandwidth, of $NEP_{ph}=0.9\times 10^{-16}W/\sqrt{Hz}$ (which converts into a Noise Equivalent Flux Density of 0.30mJy/beam). To this, in order to estimate the real sensitivity of MISTRAL, we have to add the atmospheric effects due to turbulence. An a-priori calculation of this effect is quite complicated because it also depends on the kind of filtering and on the scanning strategy: the best option would be directly measuring the atmosphere with MISTRAL. This will happen during the commissioning phase, but it is possible to give an early estimate by using experimental data taken at different frequencies. We have SRT data in the K-band (at 22GHz) and we can assume that the ratio between atmospheric fluctuations is equal to the squared ratio of the opacities. In this way, we can infer the atmospheric fluctuation noise at 90GHz using atmospheric modelling softwares such as $am$\footnote{\url{https://www.cfa.harvard.edu/~spaine/am/}}. This increases the atmospheric noise estimation a factor 10 with respect to the photon noise level even in the best 5$\%$ percentile resulting in a noise Noise Equivalent Flux Density of $3mJy/beam$. The final result is that MISTRAL detectors will be dominated by atmospheric noise rather than by electric noise, as desired\cite{Isopi2021}. 

\subsection{Scanning strategy}

The final noise level of our maps will depend on the detectors sensitivity, on the atmospheric contamination but also on the scanning strategy and the filtering of the data. We can clearly clean observations very aggressively at the cost of loosing large angular scales on sky. Among the many scientific goals of MISTRAL, the possibility of retrieving large angular scales at the level of its field of view (i.e. 4 arcmin) and even more would be crucial for the study of galaxy clusters and, for instance, for the search for the Cosmic Web in close-by clusters. Scanning strategy drives also the sensitivity as we can more or less efficiently remove atmosphere emission and its fluctuations

\begin{figure}[h]
    \centering
    \includegraphics[width=140mm]{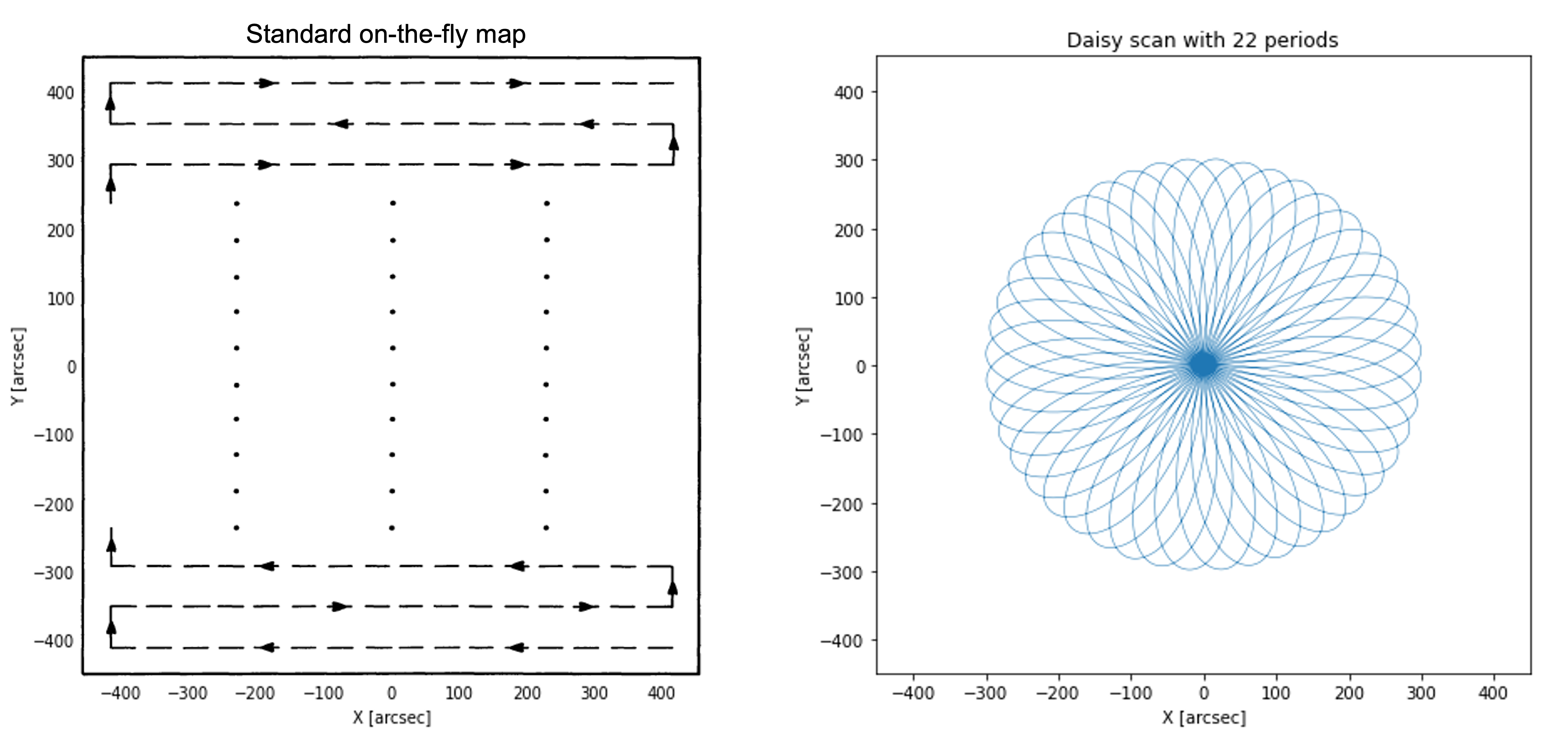}
    \caption{Left: on-the-fly map scanning strategy already implemented at the SRT. Right: simulation of 22 periods dasy scanning over a 10 arcmin sky area.}
    \label{fig:scan}
\end{figure}

At the SRT it is already in use an on-the-fly map scanning strategy which allows orthogonal scans in R.A./Dec or Azimuth/Elevation each composed by parallel sub-scans. In addition, we are working on the scanning strategy based on Lissajou daisy scanning already in use on different Telescopes such as, for instance, the GBT\footnote{\url{https://www.gb.nrao.edu/scienceDocs/GBTog.pdf}} which has the advantage to reduce to the minimum the overhead due to Telescope inversion and optimize the integration time on the center of an observed field  (see figure \ref{fig:scan}). This kind of scanning strategy can easily be analysed with Fourier space filtering of (typically) 0.1Hz (cut-on) and 7.5Hz (cut-off frequency).

\section{Science case}

MISTRAL will be a facility instrument. Thus, it will be open to the scientific community to decide what kind of scientific output it can achieve and propose the observations to the Time Allocation Committee of the Sardinia Radio Telescope. Nevertheless, we will list in the following a few science cases including those that we think are the most interesting. 

Protoplanetary discs millimetric measurements in star forming regions allow to break the degeneracy present in InfraRed (IR) measurements due to the optically thick nature of the hot inner disc. Measuring the SED at mm wavelengths with high angular resolution (i.e. $\simeq$ 10 arcsec) can add information to planetary formation theories \cite{Petersen_2019} . 
Star formation in molecular clouds with high enough angular resolution allow to distinguish starless cores with respect to those hosting protostars. This is for instance what was attempted with the Large Millimeter Telescope\footnote{\url{http://lmtgtm.org/}}$^{,}$\cite{Sokol}. 
Continuum resolved galaxies observations can give information about morphology and radial profiles (gas column profiles, dust temperature profiles)\cite{Wall}. Spatially resolved high-z radio galaxies can provide info about cold dust re-emission\cite{Humphrey}.

\begin{figure}[h]
    \centering
    \includegraphics[width=\textwidth]{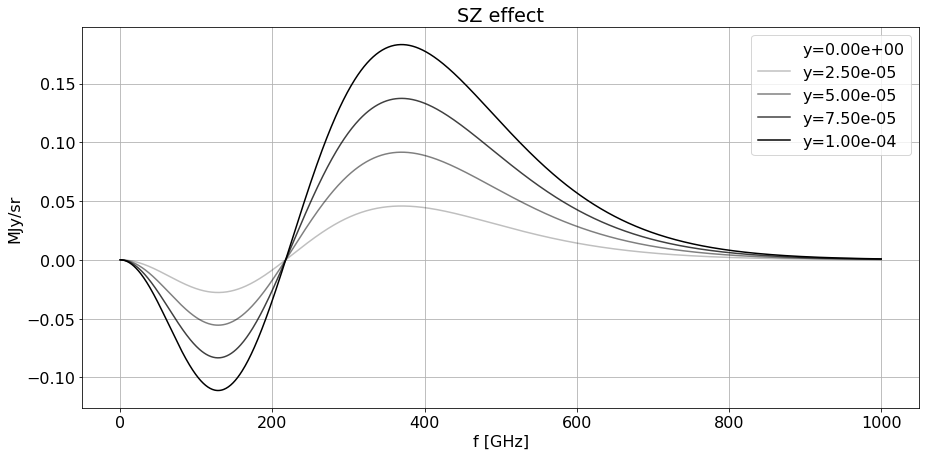}
    \caption{The SZ effect with different Comptonization parameters.}
    \label{fig:SZE1}
\end{figure}

When the CMB photons scatter off a hot electron gas, for example in galaxy clusters, they undergo inverse Compton scattering which is visible in the frequency spectrum of the CMB. The resulting distorted spectrum is thus:

\begin{equation}
\frac{\Delta I(x)}{I_0} = y \frac{x^4e^x}{(e^x-1)^2}\left(x\coth \frac{x}{2}-4 \right)= yg(x) 
\end{equation}
where: $I_0 = \frac{2h}{c^2}\left(\frac{k_b T_{CMB}}{h} \right)^3 $, $T_{CMB}$ is the CMB temperature, $x=\frac{h\nu}{k_{b}T_{CMB}}$ is the adimensional frequency, and $y=\int{n_{e}\sigma_{T}\frac{k_{B}T_{e}}{m_{e}c^{2}}}$ is the Comptonization parameter, the integral along the line of sight $dl$ of the electron density $n_{e}$ multiplied by the electron temperature $T_{e}$ (see figure \ref{fig:SZE1}). This is the Sunyaev Zel'dovich (SZ) effect\cite{SZ_1972,Birkinshaw_1999} and can be used to study the physics of Galaxy clusters. In fact, galaxy clusters can experience a wide variety of situation including collisions, merging, and non-relaxed situation. In addition, not only galaxy clusters can host relativistic electrons.  Hydrodynamical simulations \cite{Cen_1999,Tuominen_2021} suggested that galaxy clusters occupy the knots of the so called Cosmic Web and Warm to Hot Intergalactic Medium (WHIM) is disposed connecting galaxy cluster forming filaments that could be seen through the SZ effect itself.

What is probably one of the most impacting expected result of MISTRAL is in fact the possibility to observe and resolve the SZ effect through galaxy clusters and surrounding medium. With respect to moderate to low angular resolution observations, high angular resolution ($\simeq$ 10 arcsec) detections can investigate non relaxed clusters, study merging clusters, understanding the self similarities of galaxy cluster, study the pressure profiles and understand the AGN feedback expected in some environments.

Understanding the self similarities of galaxy clusters, their identical appearance or scaled properties regardless of their mass or distance is crucial. Self-similarityseems demonstrated but this is not always the case. Galaxy clusters in fact interact, collide, merge. This brings to important degeneracies and deviation from specific models. High angular resolution SZ effect measurements allow to solve the density vs temperature profile degeneracy together with X-ray information especially for high z clusters. In order to understand the physics in different environments, it is very important to study the Intra Cluster Medium (ICM) thermodynamics, including the impact of feedback, bulk and turbulent motions, substructures and cluster asphericity.

Sometimes, clusters are assumed to follow the $\beta$ profile\cite{cavaliere_fusco_fermiano_1976}, the generalized Navarro Frank and Withe (gNFW)\cite{nfwprofile1996} profile, or the Universal model\cite{arnaud2009}. High angular resolution SZ measurements allow to disentangle between models and to identify the most appropriate ones. Even relaxed clusters experience pressure fluctuations and compressions of the ICM. The study of pressure power spectrum allows to separate relaxed vs non-relaxed clusters where larger perturbations increase in the outer shells of the clusters (where SZ is more sensitive with respect to X-ray). Resolved SZ measurements allow to infer cluster masses from integrated Y mass (Y-M) relations. The Y-M relation is sensitive to the cluster astrophysics, radiative cooling, star formation, energy injection from stars and AGN feedback. In general, it allows to study of the pressure profiles and fluctuations of the ICM. This, again, can allow to understanding the self-similarity of galaxy clusters.

\begin{figure}[h]
    \centering
    \includegraphics[width=0.75\textwidth]{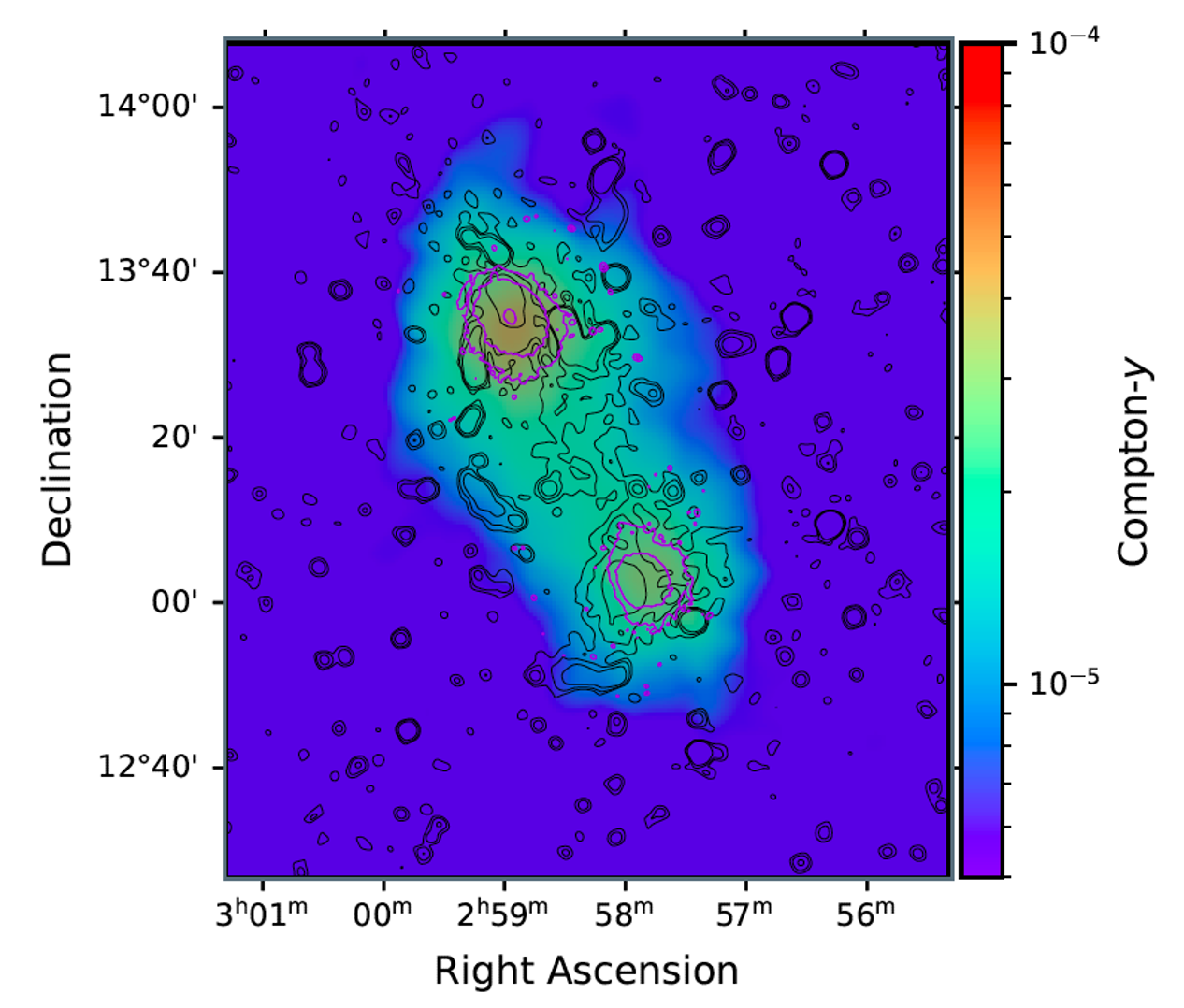}
    \caption{Planck PR2 MILCA Compton-$y$ map of Abell 401 (north-east) and Abell 399 (south-west). Black contour level are from the LOFAR map\cite{Govoni2019} at 0.0035, 0.008 and 0.02 Jy. Magenta contour level at 0.00008, 0.0015 and 0.001 Counts/s refer to the XMM-Newton map of the system\cite{Murgia2010}. The figure shows a good agreement between the radio and SZ signal both in the clusters and in the intracluster region, while the X-ray emission is appreciable only in correspondence with the two clusters.}
    \label{fig:401399}
\end{figure}

In addition, point sources can be a contamination and need to be correctly subtracted not to bias the estimate of the y-parameter in a galaxy cluster. This is for instance what was achieved with MUSTANG2 at the GBT\cite{Dicker}. 

Galaxy clusters experience hierarchical growth through mergers. Pre-mergers clusters should be connected through the Cosmic Web and simulations endorse this\cite{Cen_1999,Tuominen_2021} (see, e.g. Figure \ref{fig:401399}). On the other hand, post mergers clusters dissipate an enormous quantity of energy with turbulence and shocks (pressure discontinuity up to mach number 3). The $Bullet$ cluster or $El$ $Gordo$ are example of this unrelaxed occurrence.

The wide majority of the SZ is in fact localized towards galaxy clusters but the baryons distribution is still an open issue for modern cosmology: half of the baryons are still missing. Magneto-hydrodynamical and hydrodynamical simulations predict that they are structured in the cosmic web, filaments, and voids in the form of Warm Hot Intergalactic Medium (WHIM)\cite{Cen_1999,Tuominen_2021}. WHIM are expected to be distributed as over-desities in filamentary structures between galaxy clusters: high angular resolution SZ measurements can detect WHIM better than X-ray under low density circumstances. This is probably the most challenging and most rewarding of the achievements that an experiment like MISTRAL can achieve.

\section{Conclusions}

High angular resolution millimetric observation are key to understand a wide variety of scientific cases including structure formation and missing baryons issue. MISTRAL is a facility millimetric camera that will study several scientific case and that will be installed and commissioned at the Sardinia Radio Telescope during 2022. MISTRAL is going to be a unique camera: agile cryostat, 408 KIDs, optics allowing 12 arcsec resolution and a 4 arcmin f.o.v. at 90GHz. Sapienza University will do the technical commissioning, will participate to the astronomical commissioning, and will release MISTRAL to the scientific community.

\section{Acknowledgments }
The Enhancement of the Sardinia Radio Telescope (SRT) for the study of the Universe at high radio frequencies is financially supported by the National Operative Program (Programma Operativo Nazionale - PON) of the Italian Ministry of University and Research Research and Innovation 2014-2020, Notice D.D. 424 of 28/02/2018 for the granting of funding aimed at strengthening research infrastructures, in implementation of the Action II.1 - Project Proposal PIR01\_00010.


\begin{thebibliography}{10}

\bibitem{dickermustang2}
S.~R. {Dicker}, P.~A.~R. {Ade}, J.~{Aguirre} {\em et~al.}, {MUSTANG2: a large
  focal plan array for the 100 meter Green Bank Telescope}, in {\em Millimeter,
  Submillimeter, and Far-Infrared Detectors and Instrumentation for Astronomy
  VII\/},  eds. W.~S. {Holland} and J.~{Zmuidzinas}, Society of Photo-Optical
  Instrumentation Engineers (SPIE) Conference Series, {\bf 9153} July 2014.

\bibitem{Catalano2018}
A.~Catalano, R.~Adam, P.~A.~R. Ade, P.~Andre', H.~Aussel, A.~Beelen, A.~Benoit,
  A.~Bideaud, N.~Billot, O.~Bourrion,  {\em et~al.}, The nika2 instrument at
  30-m iram telescope: Performance and results, {\em Journal of Low Temperature
  Physics} {\bf 193}, 916 (Mar 2018).

\bibitem{Adam2018}
R.~Adam {\em et~al.}, The nika2 large-field-of-view millimetre continuum camera
  for the 30 m iram telescope, {\em A\&A} {\bf 609}, p.
  A115 (January 2018).

\bibitem{Perotto2020}
L.~Perotto {\em et~al.}, Calibration and performance of the nika2 camera at the
  iram 30-m telescope, {\em A\&A} {\bf 637}, p. A71 (May
  2020).

\bibitem{Bolli_2015_commissioning}
P.~Bolli, A.~Orlati, L.~Stringhetti {\em et~al.}, Sardinia radio telescope:
  General description, technical commissioning and first light, {\em Journal of
  Astronomical Instrumentation} {\bf 04}, p. 1550008 (Dec 2015).

\bibitem{Nasir_2011}
F.~T. Nasir, F.~Buffa and G.~L. Deiana, Characterization of the atmosphere
  above a site for millimeter wave astronomy, {\em Experimental Astronomy} {\bf
  29}, 207 (Apr 2011).

\bibitem{Govoni2021}
F.~Govoni {\em et~al.}, The high-frequency upgrade of the sardinia radio
  telescope, {\em 2021 XXXIVth General Assembly and Scientific Symposium of the
  International Union of Radio Science} {\bf }, 1 - 2021).

\bibitem{coppolecchia_LTD19}
A.~{Coppolecchia}, E.~{Battistelli}, {Masi} {\em et~al.}, {\emph{Pulse tube
  cooler with $>100$m flexible lines optimized for operation of cryogenic
  detector arrays at large radiotelescopes}} {\em 19$^{th}$ International
  Workshop on Low Temperature Detectors} in press.

\bibitem{Paiella2016}
A.~Paiella {\em et~al.}, Development of lumped element kinetic inductance
  detectors for the w-band, {\em Journal of Low Temperature Physics} {\bf 184},
  97 (July 2016).

\bibitem{Coppolecchia2020}
A.~Coppolecchia {\em et~al.}, W-band lumped element kinetic inductance detector
  array for large ground-based telescopes, {\em Journal of Low Temperature
  Physics} {\bf 199}, 130 (April 2020).

\bibitem{Paiella_2019_olimpo}
A.~Paiella, E.~S. Battistelli, M.~G. Castellano {\em et~al.}, Kinetic
  inductance detectors and readout electronics for the olimpo experiment, {\em
  Journal of Physics: Conference Series} {\bf 1182}, p. 012005 (Feb 2019).

\bibitem{paiella_LTD19}
A.~Paiella {\em et~al.}, {MISTRAL and its KIDs} {19th International Workshop
  on Low Temperature Detectors (LTD19)}, in press.

\bibitem{paiella_wolte14}
A.~{Paiella}, E.~S. { Battistelli}, P.~{ de Bernardis} {\em et~al.}, {The
  cryogenic detectors of MISTRAL} {\em 14$^{th}$ Workshop On Low Temperature
  Electronics} April 2021.

\bibitem{Isopi2021}
G.~Isopi, The mistral receiver: read-out electronics and observational
  strategies, {\em MASTER degree thesis} {\bf } (October 2021).

\bibitem{Petersen_2019}
M.~S. Petersen, R.~A. Gutermuth, E.~Nagel, G.~W. Wilson and J.~Lane, Early
  science with the large millimetre telescope: new mm-wave detections of
  circumstellar discs in ic 348 from lmt/aztec, {\em MNRAS} {\bf 488}, 1462
  (Sep 2019).

\bibitem{Sokol}
A.~D. Sokol, R.~A. Gutermuth, R.~Pokhrel {\em et~al.}, Early science with the
  large millimetre telescope: An lmt/aztec 1.1 mm survey of dense cores in the
  monoceros r2 giant molecular cloud, {\em MNRAS} {\bf 483}, 407 (Feb 2019).

\bibitem{Wall}
W.~F. Wall, I.~Puerari, R.~Tilanus {\em et~al.}, Continuum observations of m 51
  and m 83 at 1.1 mm with aztec, {\em MNRAS} {\bf 459}, 1440 (Jun 2016).

\bibitem{Humphrey}
A.~Humphrey, M.~Zeballos, I.~Aretxaga {\em et~al.}, Aztec 1.1-mm images of 16
  radio galaxies at 0.5 $<$ z $<$ 5.2 and a quasar at z= 6.3, {\em MNRAS} {\bf
  418}, 74 (Nov 2011).

\bibitem{SZ_1972}
R.~A. Sunyaev and Y.~B. Zeldovich, The observations of relic radiation as a
  test of the nature of x-ray radiation from the clusters of galaxies, {\em
  Comments on Astrophysics and Space Physics}  (Nov 1972).

\bibitem{Birkinshaw_1999}
M.~Birkinshaw, The sunyaev-zel'dovich effect, {\em Physics Reports} {\bf 310},
  p. 97?195 (Mar 1999).

\bibitem{Cen_1999}
R.~Cen and J.~P. Ostriker, Where are the baryons?, {\em The Astrophysical
  Journal} {\bf 514}, 1 (Mar 1999).

\bibitem{Tuominen_2021}
T.~Tuominen, J.~Nevalainen, E.~Tempel {\em et~al.}, An eagle view of the
  missing baryons, {\em A\&A} {\bf 646}, p. A156 (Feb
  2021).

\bibitem{cavaliere_fusco_fermiano_1976}
A.~{Cavaliere} and R.~{Fusco-Femiano}, {Reprint of 1976A\&A....49..137C. X-rays
  from hot plasma in clusters of galaxies.}, {\em aap} {\bf 500}, 95 (May
  1976).

\bibitem{nfwprofile1996}
J.~F. Navarro, C.~S. Frenk and S.~D.~M. White, The structure of cold dark
  matter halos, {\em The Astrophysical Journal} {\bf 462}, p. 563 (May 1996).

\bibitem{arnaud2009}
M.~{Arnaud}, {The {\ensuremath{\beta}}-model of the intracluster medium.
  Commentary on: Cavaliere A. and Fusco-Femiano R., 1976, A\&A, 49, 137}, {\em
  aap} {\bf 500}, 103 (June 2009).

\bibitem{Govoni2019}
F.~Govoni {\em et~al.}, A radio ridge connecting two galaxy clusters in a
  filament of the cosmic web, {\em Science} {\bf 364}, 981 (June 2019).

\bibitem{Murgia2010}
M.~Murgia {\em et~al.}, A double radio halo in the close pair of galaxy
  clusters abell 399 and abell 401, {\em A\&A} {\bf 509},
  p. A86 (January 2010).

\bibitem{Dicker}
S.~R. Dicker, E.~S. Battistelli, T.~Bhandarkar {\em et~al.}, Observations of
  compact sources in galaxy clusters using mustang2, {\em MNRAS} {\bf 508}, 2, 2600 (Dec 2021).

\end{thebibliography}
\end{document}